\documentclass[aps,prx,reprint,groupedaddress]{revtex4-1}
\usepackage{latexsym,epsfig,bm,times,psfrag,subfigure}
\usepackage{bbm}
\usepackage{color}
\pdfoutput=1

\begin{document}

\title{Construction of Many-Body Eigenstates with Displacement Transformations}

\author{Miguel Ortu\~{n}o}
\author{Andres M. Somoza}
\affiliation{Departamento de F\'{i}sica - CIOyN, Universidad de Murcia, Murcia 30.071, Spain}
\author{Louk Rademaker}
\affiliation{Department of Theoretical Physics, University of Geneva, 1211 Geneva, Switzerland}

\begin{abstract}
Many-body eigenstates beyond the gaussian approximation can be constructed in terms of local integrals of motion (IOM), although their actual computation has been until now a daunting task.
 We present a new practical computation of IOMS based on displacement transformations.
 It represents a general and systematic way to extend Hartree-Fock and configuration interaction theories to higher order.
 Our method combines minimization of energy and energy variance of a reference state with exact diagonalization. We show that our implementation is able to perform ground state calculations with high precision for relatively large systems.  Since it keeps track of the IMO's forming a reference state, our method is particularly efficient dealing with excited states, both in accuracy and the number of different states that can be constructed. 
 
\end{abstract}
\maketitle

\section{Introduction}

Integrals of motion (IOM) have gained a renewed interest in the context of disordered many-body systems. In the non-interacting Anderson insulator \cite{Anderson:1958fz} in $d=1,2$ dimensions the single particle wavefunctions are exponentially localized for any degree of disorder. In the presence of  interactions particles will remain localized, which is known as many-body localization (MBL) for strong disorder\cite{BaskoAA,Nandkishore:2015kt,Znidaric:2008cr,Pal:2010gr,Bardarson:2012gc,DeLuca:2013ba,Abanin2017,Alet}.  
It was realized that MBL can be understood through the existence of an extensive number of exponentially localized IOMs \cite{Chandran,Huse:2014co,Serbyn:2013cl}. Inevitably, this observation led to a rush of new methods to compute the IOMs in the MBL-phase,\cite{2015NuPhB.891..420R,Imbrie2016,YouQiXu2016,2016arXiv160707884P,
2016arXiv160801328G,
2016arXiv160803296O,
2016arXiv160300440I,Friesdorf:2015dd,2016arXiv160609509H,ImbrieReview,Peng2019,Varma2019,Mierzejewski}.
We published our own method based on displacement transformations of the Hamiltonian written in second quantization, which was one of the first practical computational approaches \cite{Rademaker:2016jf, ROS17}.

The presence of these localized IOMs in the fully many-body localized phases prevents thermalization. The question of whether a many-body quantum system thermalizes has been cast into the Eigenstate Thermalization Hypothesis (ETH):\cite{Deutsch:1991ju,1996JPhA...29L..75S,Srednicki:1994dl,Rigol:2011bf} the expectation value of any local observable in an eigenstate with a given energy density is equal to its expectation value in the Gibbs ensemble with corresponding temperature. If, however, there exist local density IOMs (that can be expressed as the sum of local operators) the corresponding thermal state will be a so-called generalized Gibbs ensemble.\cite{Rigol:2006df} It has been shown that any finite-ranged translationally invariant Hamiltonian will thermalize towards their corresponding generalized Gibbs ensemble.\cite{2015arXiv151203713D} Therefore, whether and how a system thermalizes is directly related to the structure of its integrals of motion.

The existence of IOMs has a deep consequence for the structure of many-body eigenstates. Written in the basis of IOMs, {\em all} many-body eigenstates are just product states defined by the occupation of each IOM. Given the basis transformation $U$ from the physical basis to the basis of IOMs, eigenstates can be written as $U |\psi_{SD} \rangle$ where $|\psi_{SD}\rangle$ is a Slater determinant of the form $c_{\beta_1}^\dag\cdots c_{\beta_N}^\dag|0\rangle$. 

Though an accurate computation of all IOMs is in practice very tedious, we propose here that one can focus on a specific reference (product) state. By only considering those displacement transformations that affect that reference state, we increase the efficiency and thus accuracy of the numerical implementation. In this work we show that this indeed works, and we find accurate results for ground states and excited states for systems up to $L=30$ sites. 
In the lowest possible order, the method can work up to much larger sizes.

White \cite{White} introduced a similar method to ours to diagonalize the Hamiltonian through what he calls canonical transformations (we will see that our displacement transformation are canonical), sharing many ingredients with our approach, although no explicit mention of the IOM and their general applicability  was made.  
Both methods deal directly with the Hamiltonian, instead of the wavefunctions, unlike most other numerical approaches. They manipulate the second quantized Hamiltonian, which is a sum of abstract operator terms with the corresponding numerical coefficient. This presents several advantages. For example, one can efficiently separate the low-energy and high-energy orbitals from the strongly correlated intermediate ones. Then, fixing the occupancy of low-energy levels to 1 and of high-energy levels to 0, one can remove them from the problem. A second advantage is that basically the same approach is used for the ground state and for excited states.

The so called flow equation method\cite{Wegner,Wegner2} and similarity renormalization\cite{Wilson} are continuous versions of unitary transformations, in which a set of differential equations is solved to implement the displacement transformations.\cite{2016arXiv160707884P,Monthus2016,Savitz,Monthus3,Monthus4,Thomson} Continuous transformations have the advantage of commuting one with each other, while discrete implementations take into account the term to be transformed at all orders in a single step.

The remainder of the paper is set up as follows. In Sec.~\ref{SecGenHam} and \ref{SecDispl} we briefly review the concept of IOMs and our method of displacement transformations. The increase in efficiency gained by focusing on a specific reference state is explained in Sec.~\ref{SecRef}. We proceed with comments on the practical implementation (Sec.~\ref{SecPrac}) and some numerical results for a one-dimensional interacting disordered system in Sec.~\ref{SecNum}. In Sec.~\ref{SecOut} we discuss possible applications and related work.


\section{General quantum Hamiltonian}
\label{SecGenHam}

We consider a general fermionic quantum Hermitian Hamiltonian, which can always be written in the form 
\begin{equation}
H = \sum_X V_X (X^\dag+X)
\label{general}\end{equation}
with the terms $X$ of the form
\begin{equation}
X=n_{\alpha_1}\cdots n_{\alpha_k} c_{\beta_1}^\dag c_{\gamma_1} c_{\beta_2}^\dag
\cdots c_{\gamma_l}
\label{general1}\end{equation}
where $c_i^\dag$ ($c_i$) is the fermionic creation (annihilation) operator of state $i$ and $n_i=c_i^\dag c_i$ is the number operator, and all $\alpha, \beta, \gamma$ are different. 
We will assume particle conservation and so the number of creation and annihilation operators will be the same in each term of the Hamiltonian.
Terms containing creation and annihilation operators are called non-diagonal, while terms composed of density operators only are called diagonal or classical.

In principle, our goal is to find a unitarity transformation $U$ that transforms the Hamiltonian $H$ into a classical Hamiltonian $\tilde{H}=U^\dag HU$, i.e., one that only contains density operators. This "diagonalized" Hamiltonian can be written in the form
\begin{equation}
	\tilde{H} = \sum_i \epsilon_\alpha \tau_\alpha  +  \sum_{\alpha,\beta} \epsilon_{\alpha,\beta} \tau_\alpha \tau_\beta+
\sum_{\alpha,\beta,\gamma} \epsilon_{\alpha,\beta,\gamma} \tau_\alpha \tau_\beta \tau_\gamma+\cdots
\end{equation}
where $\tau_\alpha= U^\dag n_\alpha U=\tilde{c}_\alpha^\dag \tilde{c}_\alpha$ are the transformed number operators and correspond to the integrals of motion (IOMs). Now $\tilde{c}_\alpha^\dag$ and $ \tilde{c}_\alpha$ are the transformed creation and annihilation operators. Any many-body eigenstate can be constructed by application of the transformed creation operators on the vacuum state $|0\rangle$,
\begin{equation}
	|\Phi \rangle= \tilde{c}_{\beta_1}^\dag\cdots \tilde{c}_{\beta_N}^\dag|0\rangle
	= U^\dag c_{\beta_1}^\dag\cdots c_{\beta_N}^\dag|0\rangle.
	\label{iom}
\end{equation}
In the basis of IOMs, therefore, all many-body eigenstates are `product states' in the transformed creation operators. This also implies that any Slater determinant of the original fermions becomes a many-body eigenstate under the unitary transformation $U$.

\section{Displacement transformations}
\label{SecDispl}

Two of us showed earlier\cite{Rademaker:2016jf} that the Hamiltonian given by Eq.\ (\ref{general}) can be diagonalized by a unitary transformation consisting of a product of displacement transformations.
A displacement transformation  is defined as 
\begin{equation}
{\cal D}_X(\lambda)=\exp\{\lambda(X^\dag-X)\},
\label{displacement}\end{equation}
 where $X$ is any given quantum term. It is trivial to show that ${\cal D}_X(\lambda)$  is equal to
\begin{equation}
{\cal D}_X(\lambda)=\mathbf{1}+\sin(\lambda) (X^\dag-X)+(\cos(\lambda) -1)(X^\dag X+XX^\dag).\label{displace}
\end{equation}
The fact that the displacement transformation can be summed in a compact way has important practical consequences, as we will see.

To start with, any operator transformed under ${\cal D}_X(\lambda)$ can be cast in a closed expression.
Defining $A\equiv X^\dag -X$ and focusing in the Hamiltonian $H$, we arrive at
\begin{eqnarray}
	 {\cal D}_X(-\lambda)H(\lambda){\cal D}_X(\lambda) &=&  H  + \sin\lambda   [H,A] 
	\nonumber\\&-&(\cos\lambda -1) HA^2+A^2H  \\
	&-&\sin\lambda(\cos\lambda -1) A^2HA-AHA^2 \nonumber\\
	&-&\sin^2\lambda   AHA +(\cos\lambda -1)^2 A^2HA^2 \nonumber
	\label{compact}\end{eqnarray}
This is in contrast with the typical infinite series expansion of the transformed operator in terms of commutators of $A$ and the operator:
\begin{eqnarray}
&&	{\cal D}_X(-\lambda)H(\lambda){\cal D}_X(\lambda) =  H  + \lambda   [H,A]\\
&&+\frac{\lambda^2}{2!} [[H,A],A]  
	+\frac{\lambda^3}{3!} [[[H,A],A],A]+\cdots \nonumber
		\label{compact2}\end{eqnarray}

Under a displacement transformation ${\cal D}_X(\lambda)$, the prefactor of the term $X^\dag+X$ in the new Hamiltonian is changed and depends on the prefactors in the old Hamiltonian of the term itself and of the density operators corresponding to the creation and annihilation operators of $X$. Both contributions can cancel each other if the strength of the transformation $\lambda$ is chosen as (see Appendix A)
\begin{equation}
\tan (2\lambda) =\frac{2V_X}{\Delta \epsilon_X}
\label{tan2}\end{equation}
where $V_X$ is the coefficient of the operator $X^\dag+X$ in the Hamiltonian
and $\Delta \epsilon_X$ is the energy difference between the two configurations involved in the transition $X$. That is, if $X$ is given by Eq.\ (\ref{general1}), then
\begin{eqnarray}
\Delta \epsilon_X&=&\epsilon_{\beta_1}+\cdots+\epsilon_{\beta_l}-
 \epsilon_{\gamma_1}-\cdots-\epsilon_{\gamma_l}\\
&+&\epsilon_{\alpha_1,\beta_1}+\cdots+\epsilon_{\alpha_1,\beta_l}-
 \epsilon_{\alpha_1,\gamma_1}-\cdots-\epsilon_{\alpha_1,\gamma_l}+\cdots\nonumber
\label{difference}\end{eqnarray}
where $\epsilon_{\alpha_1,\beta_1}$ is the coefficient of the term $n_{\alpha_1} n_{\beta_1}$ in the Hamiltonian.

After transforming the Hamiltonian under ${\cal D}_X(\lambda)$, with $\lambda$ given by Eq.\ (\ref{tan2}), the operator $X$ only remains in the classical combination
\begin{equation}
V_X \tan (\lambda) (X^\dag X-XX^\dag)
\end{equation}

It is interesting to note that the condition of canceling out the quantum term is equivalent to extremizing (maximizing/minimizing) the coefficient of the classical term, since the derivative with respect to $\lambda$ of its contributions is equal zero when Eq.\ (\ref{tan2}) is satisfied (see Appendix A). This is just a consequence of the conservation of the trace of $H^2$ under the displacement transformations.

One can calculate any observable by constructing the corresponding operator and transforming it with the same displacement transformations that diagonalize the Hamiltonian.
As the many-body states take a simple form in this new basis, Eq.\ (\ref{iom}), it is easy to compute the expectation value and the variance of the operator.

One can in principle diagonalize the interacting Hamiltonian by the application of successive displacement transformations, starting with those involving operators of the form $c_j^\dag c_i$ and continuing with higher (larger number of terms) order operators. In practice this method generates and exponentially large number of terms, as was shown in our earlier work\cite{Rademaker:2016jf,ROS17}. As we will show in the following sections, it is more efficient to concentrate on certain expectation values with respect to a given state. This allows us to study with more accuracy certain parts of the spectrum.

We pause here to note that our displacement transformations are canonical transformations, as defined in (\ref{displace}).
A transformation is canonical if  when applied to the creation $c^\dagger_i$  and annihilation $c_j$ operators preserves their anticommutation relations.
Our displacement transformations ${\cal D}_X$  are indeed canonical transformations since
\begin{equation}
	\{{\cal D}_X c^\dagger_i {\cal D}_{-X}, {\cal D}_X c_j {\cal D}_{-X}\}=
	\{c^\dagger_i,c_j\}=\delta_{i,j}.
	\label{comm}\end{equation}
In fact our displacement transformations are the main type of canonical transformation considered in Ref.\ \onlinecite{White}.

\section{Reference States}
\label{SecRef}

\subsection{Energy minimization}

Since the diagonalization of the full Hamiltonian is a huge task, it is more efficient to focus on a given state  $|\Phi_0 \rangle$ and perform only those transformations that affect the expectation value or the variance of the 
Hamiltonian with respect to this state. This is the core novel development of this work.

Let us begin by choosing transitions $X$ and their corresponding $\lambda$ so that  the displacement transformation defined by them  minimizes (or maximizes) the expectation value of the  energy 
\begin{equation}
\frac{{\rm d} \langle \Phi_0|{\cal D}_X^\dag(\lambda)H{\cal D}_X(\lambda)|\Phi_0 \rangle}{\rm d\lambda}\equiv \frac{{\rm d} \langle H(\lambda) \rangle}{\rm d\lambda}=0
\label{minimum}\end{equation}
where  our state $|\Phi_0 \rangle$ is  of the form $|\Phi_0 \rangle=c_{\beta_1}^\dag\cdots c_{\beta_N}^\dag|0\rangle$, where $|0\rangle$ is the vacuum and the set of  indices $\beta_i$ is fixed and define our state. 
The creation and annihilation operators are self-consistently redefined at each step (at each displacement transformation). Strictly speaking we minimize the energy if we are considering the ground state, while we find a local extremum of the energy if we are dealing with an excited state.
If we are interested in the ground state, $|\Phi_0 \rangle$ is our running estimate of it and the indices $\beta_i$, instead of being fixed, have to be chosen to minimize the expectation value of the energy.

Taking expectation values  with respect to state $|\Phi_0 \rangle$ in (\ref{compact}) we arrive at
\begin{eqnarray}
	\langle H(\lambda) \rangle&=& \langle H \rangle + \sin\lambda  \langle [H,A] \rangle
	\nonumber\\&-&(\cos\lambda -1)\langle HA^2+A^2H \rangle\\
	&-&\sin\lambda(\cos\lambda -1)\langle A^2HA-AHA^2 \rangle \nonumber\\
	&-&\sin^2\lambda  \langle AHA \rangle+(\cos\lambda -1)^2\langle A^2HA^2 \rangle\nonumber
	\label{expect}\end{eqnarray}
In order to have $A|\Phi_0 \rangle\ne 0$, all density operators in $X$ must correspond to occupied states in $|\Phi_0 \rangle$, while creation operators must have the opposite occupancy than annihilation operators in $|\Phi_0 \rangle$.
Displacement transformations of operators $X$ not satisfying these two conditions do not have to be considered as long as the state $|\Phi_0 \rangle$ is involved. This is at the heart of the increase in efficiency of our implementation.

For the transformations affecting the expectation value of $H$, we have $ \langle A^2 \rangle =-1$ and substituting in (\ref{expect}) we get
\begin{eqnarray}
\langle H(\lambda) \rangle&=& \langle H \rangle + \sin\lambda\cos\lambda  \langle [H,A] \rangle
\nonumber\\
&-&\sin^2\lambda  \left(\langle AHA \rangle+\langle H \rangle\right)
\label{expect2}\end{eqnarray}

We have the freedom to choose without loss of generality the operator $X$ (as opposed to $X^\dagger$) such that $X|\Phi_0 \rangle\ne 0$, and we will denote the resulting state as
 $|\Phi_X \rangle\equiv X|\Phi_0 \rangle$.
Then
\begin{equation}
-\langle AHA \rangle-\langle H \rangle=\langle\Phi_X| H|\Phi_X \rangle-\langle\Phi_0| H|\Phi_0 \rangle\equiv \Delta E_X
\label{energy}\end{equation}
This quantity is the energy difference between the states $|\Phi_X \rangle$ and $|\Phi_0 \rangle$ and so only depends on the classical terms of $H$.
Let us define the expectation values of the Hamiltonian multiplied by different operators $Y$ and $Z$ as
\begin{equation}
V_{Y,Z}\equiv \langle\Phi_0|Y^\dagger H Z|\Phi_0 \rangle
\label{vxy}\end{equation}
From now on we will consider a real Hamiltonian, and so 
\begin{equation}
\langle\Phi_0| [H,A]|\Phi_0 \rangle=-2 V_{X,{\bf 1}}
\label{conm}\end{equation}
where ${\bf 1}$ denotes that the unity operator is acting on $|\Phi_0 \rangle$.
We note that all terms of the Hamiltonian with the same quantum part as $X$ and with  density operators of states occupied in $|\Phi_0 \rangle$ will contribute to $V_{X,{\bf 1}}$.
With the previous definitions, $\langle H(\lambda) \rangle$ becomes
\begin{equation}
\langle H(\lambda) \rangle= \langle H \rangle - \sin(2\lambda)  V_{X,{\bf 1}}
+\sin^2\lambda   \Delta E_X
\label{expect2a}\end{equation}
And the value of $\lambda$ minimizing (or maximizing) $\langle H(\lambda) \rangle$ is
\begin{equation}
\tan 2\lambda_X =\frac{2V_{X,{\bf 1}}}{\Delta E_X}.
\label{tan2a}\end{equation}

This equation has two solutions for $\lambda_X$ differing by $\pi/2$. They correspond to the two possible ways to associate the bonding and antibonding states with the original states.
In order to maximize the overlap with the initial state, it is convenient to choose the solution satisfying $|\lambda|<\pi/4$.

One can transform $H$ reiteratively with different displacement transformations up to a given order until all coefficients $V_{X,1}$ are smaller than a given cutoff.

Considering only second order displacement transformations, our method is equivalent to the standard Hartree-Fock approximation.
Two important advantages of our method is that it can naturally be extended by considering higher order transformations and that it can be applied efficiently to excited states, as we will see.

\subsection{Energy variance minimization}

The previous scheme of getting extreme values for the energy works fairly well for ground state calculations and in some cases for excited states, but not in general.     
Considering the MBL problem, for example, energy minimization (maximization) is efficient in the localized regime, but not so in the extended regime, where it does not converge for some states, that end up oscillating in a cyclic way between different  configurations. The action of a given transformation is undone by the action of two other transformations maximally overlapped with the first one. The problem arises because we are not minimizing any function. The energy is an extreme for all eigenvalues, but not necessarily a minimum, except for the ground state.
We can get around this problem by minimizing the variance of the energy
\begin{equation}
\sigma^2\equiv \langle \Phi_0|H^2|\Phi_0 \rangle-\langle \Phi_0|H|\Phi_0 \rangle ^2
\label{quantum3}\end{equation}

This procedure is practical if we are able to calculate the value of $\lambda$ that minimizes $\sigma^2$ without having to obtain the operator $H^2$ explicitly.
Under a displacement transformation $\langle H^2(\lambda)\rangle$ is given, in analogy with Eq.\ (\ref{expect2a}), by
\begin{eqnarray}
\langle H^2(\lambda) \rangle&=& \langle H^2 \rangle - \sin(2\lambda)  \langle \Phi_X | H^2 |\Phi_0 \rangle
\nonumber\\
&+&\sin^2\lambda  \left(\langle  \Phi_X |H^2 | \Phi_X \rangle-\langle  \Phi_0 |H^2 | \Phi_0 \rangle\right)
\label{expect3}\end{eqnarray}
Inserting the identity operator $\sum_n |\Phi_n \rangle \langle\Phi_n|$, where $|\Phi_n\rangle$ are a basis of the many-body space, which can be obtained by applying all possible combinations of $N/2$ creation operators, we can rewrite $ \langle \Phi_X | H^2 |\Phi_0 \rangle$ as
\begin{eqnarray}
	&&\langle \Phi_X | H^2 |\Phi_0 \rangle=\sum_n \langle \Phi_X | H|\Phi_n \rangle \langle\Phi_n|H |\Phi_0 \rangle\\
	&=&V_{X,{\bf 1}}
  \left(E_X+E_0\right)+\sum_{Z_n\ne {\bf 1},X} V_{X,Z_n}V_{Z_n,{\bf 1}}\nonumber
\label{inter}\end{eqnarray}
where  $Z_n$ is the operator that takes the system from $|\Phi_0\rangle$ to $|\Phi_n\rangle=Z_n|\Phi_0\rangle$ and we can assume that it does not contain density operators.
With a similar procedure, we get
\begin{eqnarray}
&&\langle  \Phi_X |H^2 | \Phi_X \rangle-\langle  \Phi_0 |H^2 | \Phi_0 \rangle=
\label{inter2}\\
&&E_X^2-E_0^2+\sum_{Z_n\ne {\bf 1},X} \left(V_{X,Z_n}^2-V_{Z_n,{\bf 1}}^2 \right)\nonumber
\end{eqnarray}

Substituting Eqns.\ (\ref{expect3}-\ref{inter2}) in (\ref{quantum3}) we arrive at
\begin{eqnarray}
	\sigma(\lambda)^2
	&=&\sigma(0)^2-\sin(2\lambda)\sum_{Z_n\ne {\bf 1},X} V_{X,Z_n}V_{Z_n,{\bf 1}}\nonumber\\
	&+&\sin^2(\lambda)\sum_{Z_n\ne {\bf 1},X} \left(V_{Z_n,{\bf 1}}^2 -V_{X,Z_n}^2\right)\label{quantum2}\\
	&-&\sin(4\lambda)\frac{\Delta E_X}{2}V_{X,{\bf 1}}
+\sin^2(2\lambda)\left( \frac{\Delta E_X^2}{4}-V_{X,{\bf 1}}^2 \right)
\nonumber
\nonumber
\end{eqnarray}
This expression can be minimized with respect to $\lambda$ numerically.
We note that the terms in the third line of Eq.\ (\ref{quantum2}) contain the same parameters that appear in the expression for the energy, Eq.\ (\ref{expect2a}), and produce the same contribution to $\sigma(\lambda)^2$ at both energy minima (differing in $\lambda_X$ by $\pi/2$). 
The terms in the first and second line of Eq.\ (\ref{quantum2}) have half the period than the other terms and, in general, unbalance the contribution to $\sigma(\lambda)^2$ in the two energy minima.
This implies that when minimizing the variance we cannot restrict $\lambda$ to values satisfying the condition $|\lambda|<\pi/4$.

When obtaining excited states, we can get extremes of the energy, but with the constraint that the energy variance never increases. In this way the procedure is stable. Once there are no displacement transformations that minimize/maximize the energy up to a given cutoff, we can minimize the energy variance, which results in a further diagonalization of the Hamiltonian not affecting directly the reference state.

\subsection{Perturbation treatment of the remaining Hamiltonian}

In practice it is necessary to establish a cut-off for the strength of the transformations $\lambda$ (or for the coefficients of the off-diagonal terms of the Hamiltonian) to be performed. We then apply perturbation theory to the remaining small off-diagonal Hamiltonian and to any other operator we may want to calculate.
However, the level spacing of many-body states is so small that direct perturbation theory becomes impractical due to the uncertainty in the energy denominators. 
A convenient alternative is to `project' the Hamiltonian onto a subspace close to the reference state $|\Phi_0\rangle$ by constructing the matrix formed by the elements $V_{Y,Z}$ where $Y$ and $Z$ are (purely quantum) operators of a given order. This matrix is then diagonalized.

If we are calculating the expectation value of some other operator $\cal O$, we construct the corresponding matrix
$\langle\Phi_0|Y^\dagger {\cal O} Z|\Phi_0 \rangle$ and rotate it with the unitary transformation $U_1$ that diagonalizes the projected Hamiltonian matrix.
The element $\langle\Phi_0| U_1^\dag{\cal O}U_1 |\Phi_0 \rangle$  is then  our estimate of the expectation value of ${\cal O}$.

This final matrix diagonalization of the `projected'  Hamiltonian is similar to the so-called configuration interaction approach, usually based on a Hartree-Fock basis.
The advantage of our procedure is that we can go naturally beyond the Hartree-Fock method by performing displacement transformations of higher order. Furthermore, using the displacement transformation we have constructed a nearly diagonal Hamiltonian, even before the final diagonalization procedure. Note that since Hartree-Fock methods construct so-called gaussian states, our method allows for the construction of non-gaussian states.\cite{Nooijen2000}

as our method controls the IOM's explicitly, it is possible to carefully select which states are included in the final diagonalization, depending on the quantities to be calculated or the nature of the problem. For example, one could selesct states closest in energy to the reference state, instead of selecting in terms of the number of excitations. Or alternatively, one could try to keep states that are more likely to affect the operator to be calculated. 

\subsection{Finite density}

To overcome as much as possible the problem of the large amount of high order terms generated by the displacement transformations, it is highly convenient to consider a finite density of particles and neglect terms according to the number of excitations, instead of the number of particles. 
We consider as the "vacuum" our reference state $|\Phi_0\rangle$, which as we have seen can be either the ground state or any excited state. 
Creation and annihilation operators for occupied states in $|\Phi_0\rangle$ are transformed into annihilation and creation operators, respectively, of the corresponding hole excitations. Operators for unoccupied states are kept the same. We keep terms containing up to a given number of these new operators.

\section{Practical implementation}
\label{SecPrac}

In order to test the viability of our method, we have applied it to the important problem of many-body localization. Consider the following interacting Hamiltonian for spinless fermions on a one-dimensional chain of length $L$,
\begin{equation}
H = \sum_i \epsilon_i n_i  +  \sum_{\langle i, j\rangle} t c_j^\dag c_i+
\frac{1}{2}\sum_{\langle i, j\rangle} U n_j n_i
\end{equation}
where $c_i^\dag$ is the creation operator on site $i$ and $n_i=c_i^\dag c_i$ is the number operator.
We consider a nearest neighbor interaction with $U=1$, which sets our unit of energy, a transfer energy $t=1/2$ (to use the same values as for spins models) and a disordered site energy
$\epsilon_i\in[-W,W]$. We assume the total number of particles is equal to half the number of sites.

In our code, each term of the Hamiltonian is represented by a real coefficient corresponding to its strength and by a long integer with every two bits indicating every site operator (creation, annihilation, density or no operator). 
Operator multiplication is a very slow process to implement numerically. To overcome this shortcoming and taking into account that a displacement transformation $\mathcal{D}_X$ only modifies the operators for the sites involved in $X$, we construct a table with the outcome of the transformation for any possible combination of operators in the sites involved in $X$. Given a general product of site operators, we extract the ones at these positions, look for the outcome in the table and insert the new operators in the original product.

For this model, the ground state is always localized, while we expect to have a delocalization transition at infinite temperature (i.e., for states chosen at random) for a critical disorder around $W=3.5$ in our units.

To evaluate the degree of localization of the eigenstates of the Hamiltonian, we calculate the variance of the density operator at site $L/2$, $n_{L/2}$. 
We apply to this operator the same transformations as to the Hamiltonian. It is always expressed in our current basis.
When the procedure is finished, it is easy to evaluate the variance of this operator with respect to our reference state $|\Phi_0 \rangle$,
\begin{equation}
\sigma_{L/2}^2=\langle\Phi_0 |n_{L/2}^2|\Phi_0 \rangle-\langle\Phi_0 |n_{L/2}|\Phi_0 \rangle^2,
\end{equation}
since the operator is already in our final basis, where $|\Phi_0 \rangle$ is a single product of creation operators acting on the vacuum.

We have performed transformations up to fourth order and we have kept terms in the Hamiltonian containing up to two particle and two hole excitations with respect to the reference state.

\subsection{Ground state}

To obtain a state as close as possible to the ground state we choose the following strategy. 
We first get rid of quantum terms of second order by performing all possible displacement transformation of this order (this is equivalent to be in the one-electron basis). 
Then we choose the set of creation operators $\{\beta_i\}$ such that the state $|\Phi_0\rangle =c_{\beta_1}^\dag\cdots c_{\beta_N}^\dag|0\rangle$  minimizes the energy of the present Hamiltonian. $|\Phi_0\rangle$ is our reference state and we calculate $V_{X,{\bf 1}}$ and $\Delta E_X$ for all possible $X$, up to a given order, and  choose the one that reduces most the energy of this state. After performing the corresponding displacement transformation, we check whether a different set of creation operators reduces further the energy of the new rotated Hamiltonian. 
This step is cyclically repeated until no displacement transformation can reduce the energy any further.
A third stage looks for transformations that reduce the energy variance, keeping in this case fixed the set of creation operators defining our state.
We finally `project' the Hamiltonian over configurations differing from the reference state by one or two electron-hole excitations and diagonalize it.

\subsection{Excited states}

The basis of IOMs provides an unambiguous way to specify a many-body eigenstate, since each can be labeled by a specific choice of the IOMs.
A great advantage of our method is that, as we will see, we can get a large proportion of the excited states of the system, many more than with other available techniques. 
To construct an excited state, we select a given set of IOMs (i.e., a set of creation operators $\{\beta_i\}$) and keep performing displacement transformations that minimize (or maximize) the expectation value of the energy with respect to this state. 
Unlike for ground state calculations, the set of creation operators defining our reference state is kept fixed.
By varying systematically the set of IOMs chosen, we can get all the excited states of the system for small enough sizes. For large sizes we choose a random set of IOM to construct a state.

There are many possible minimization strategies and which is the most suitable may depend on the quantity to be studied. Here we have concentrated in trying to obtain as many different excited states as possible, i.e., to avoid that states end up collapsing in the same final state.
We found that a good strategy to this end is to perform first the transformations that most increase (or reduce) the energy, with the constraint that $|\lambda|<\pi/4$ and provided that they do not increase the variance of the energy $\sigma^2$.
Once there are no transitions changing the energy, we reduce the energy variance as much as possible.
Reducing the variance from the beginning tends to effectively interchange the IOM indices and may change the reference state, resulting in a bias in our choice of excited states.

\section{Numerical results}
\label{SecNum}

\subsection{Ground state}

\begin{figure}[h]
	\includegraphics[width=.45\textwidth]{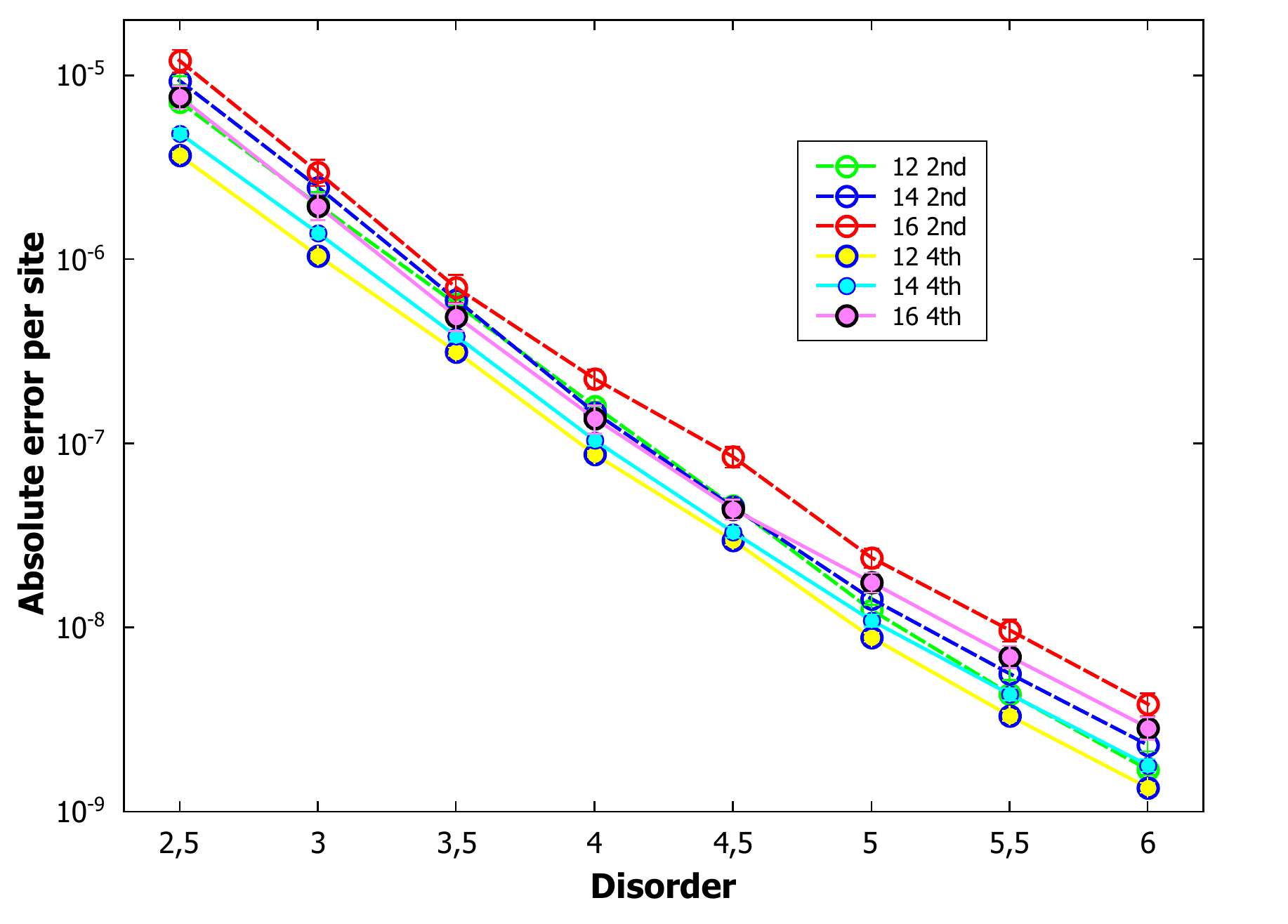}
	\caption{Absolute error of the ground state energy per site as a function of disorder for second order (empty symbols and dashed lines) and for fourth order calculations (solid symbols and continuous lines), for the system sizes indicated in the legend of the figure.}\label{error}
\end{figure}

Let us first focus on the error of our estimate of the ground state energy  as compared with results from exact diagonalization. In Fig.\ \ref{error} we present the geometrical mean of the absolute error in the ground state energy per site as a function of disorder for several system sizes and two implementations of the method.
Empty symbols and dashed lines correspond to results for 2nd order calculations, while solid symbols and continuous curves to results for transformations up to fourth order. In both cases the residual Hamiltonian was projected over configurations differing from the ground state by one or two electron-hole excitations and then diagonalized.
System sizes are  12, 14 and 16. In all cases, the cutoff strength for $\lambda$ was $10^{-3}$. 

We see in Fig.\ \ref{error} that the errors are very small for the (small) system sizes that we can diagonalize exactly. The error increases with system sizes, but does so relatively slowly.
Results for second and forth order are almost comparable, probably due to the final diagonalization performed, which for these system sizes can deal with all one and two particle excitations. 

\begin{figure}[h]
	\includegraphics[width=.45\textwidth]{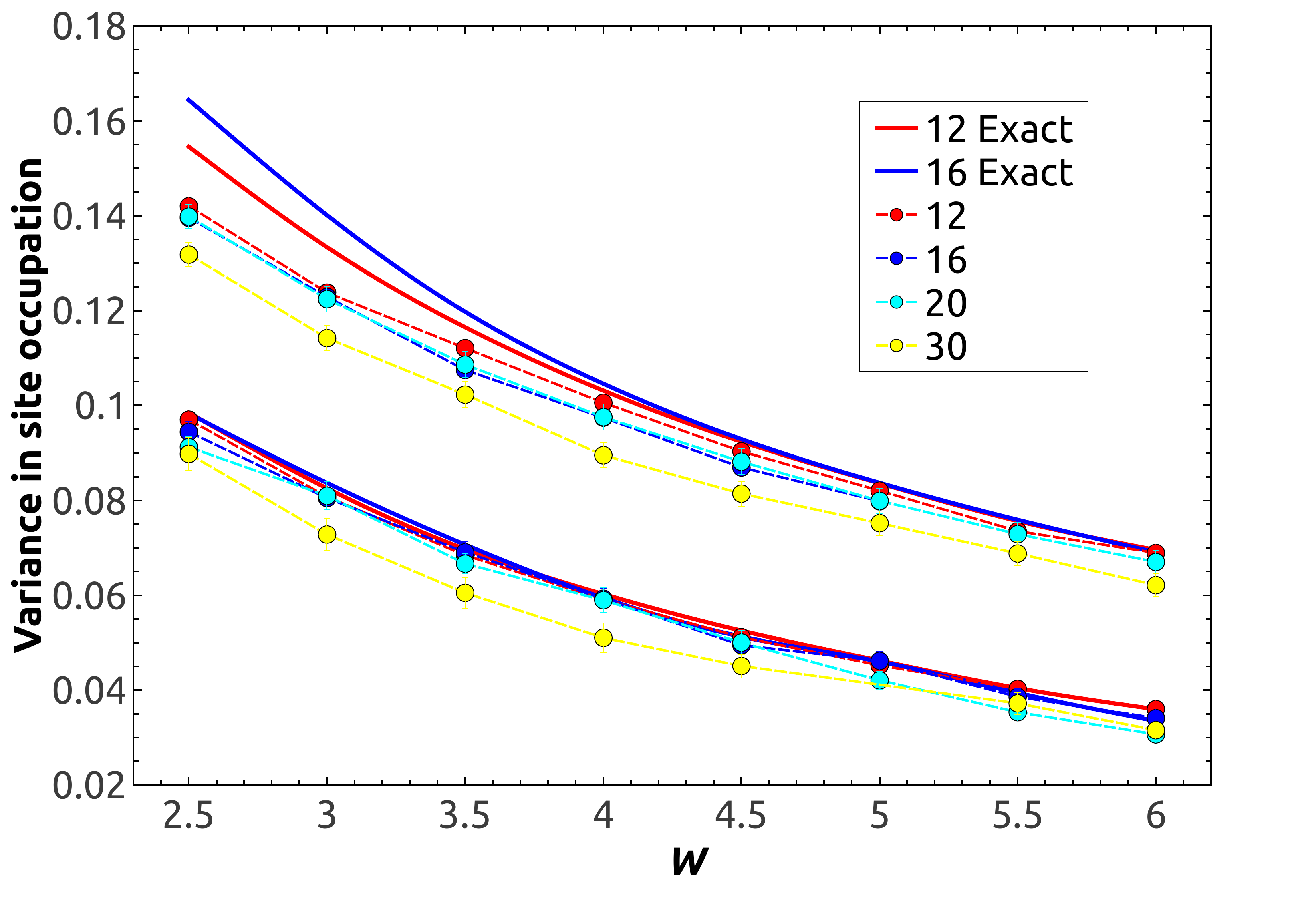}
	\caption{Variance in the occupation of a site at position $L/2$  as a function of disorder for our method (symbols and dashed lines) and for exact diagonalization (thick lines). The lower set of curves correspond to ground state calculations and includes transformations up to  fourth order, while the upper set to excited states calculations including transformations up to second order only. The system sizes are $L=12$ (black), 16 (blue),  20 (purple) and 30 (turquoise).}\label{variance}
\end{figure}

Next, we studied the variance of the occupation of the site at position $L/2$ in order to test the accuracy of our method for other quantities, beside the energy, and to prove the applicability of the method to relatively large system sizes. 
The value of the variance should be 0 in the strongly localized limit and 1/4 in the completely extended case, since we are considering 1/2 occupation.
In Fig.\ \ref{variance} we plot this variance as a function of disorder for several system sizes ranging from 12 to 30. Up to size 16, exact results are also plotted (thick lines).
Results for ground state calculations, in which we focus for the moment, correspond to the lower set of curves. Our calculation has been done up to 4th order with diagonalization of the final Hamiltonian projected over one and two particle excitations. For large system sizes we only include the first 8000 configurations ordered according to the ratio $|V_X/\Delta E_X|$, and this is the most important limiting factor at the moment.
As expected for the ground state, there is no sign of the many-body delocalization transition around $W\approx 3.5$.

The results from exact diagonalization indicate that for the ground state the variance of the number occupation is basically independent of system size. This allows us to estimate the error in this quantity for system sizes that cannot be solved exactly.
One can appreciate that our results for small system sizes are in excellent agreement with exact results, and for large system sizes we have a small deviation towards lower values of the variance.
For ground state calculations, our procedure can deal with system sizes much larger than exact diagonalization procedures and the quality of the results is quite acceptable.

The CPU time per sample for the present implementation of the algorithm up to fourth order transformations grows as $L^{10}$, implying a present practical limit at approximately $L=30$.

\subsection{Excited states}

A crucial quantity when studying excited states is the ratio between the number of states generated and the total number of states, which we will call {\em success ratio}. 
We try to generate all states by systematically varying the occupation number of the different IOMs keeping the number of particles equal to half the number of sites. Nevertheless, one can expect that sometimes the algorithm starting from different states ends up in the same final state, since the evolving dynamic is not exact and some states may be more stable than others under this algorithm.
We want to estimate  the success ratio of our procedure.
Once the level spacing is of the order of the energy uncertainty, it is difficult to calculate this quantity. 
We have designed a method to estimate the success ratio through the level statistics, as we explain in what follows.

\begin{figure*}[t]
	\includegraphics[width=.32\textwidth,height=4cm]{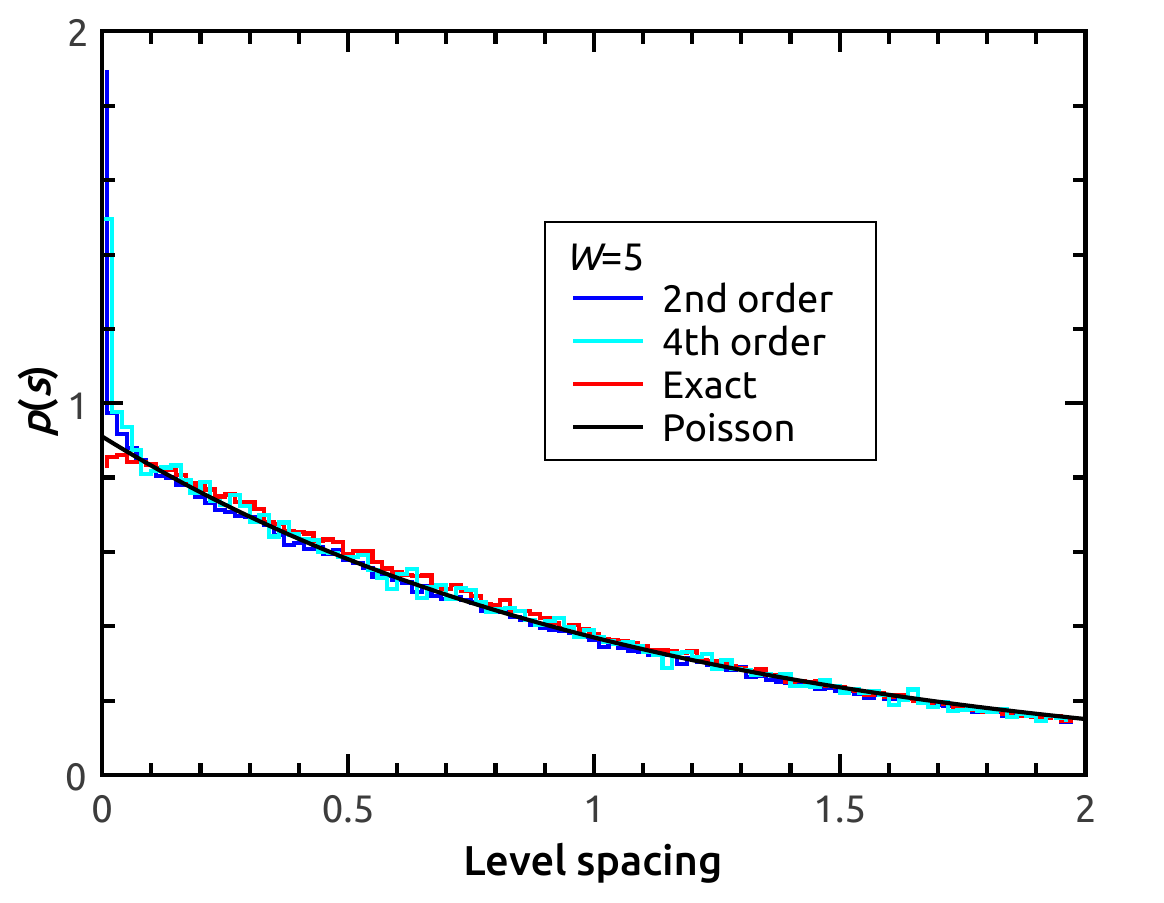}
	\includegraphics[width=.32\textwidth,height=4cm]{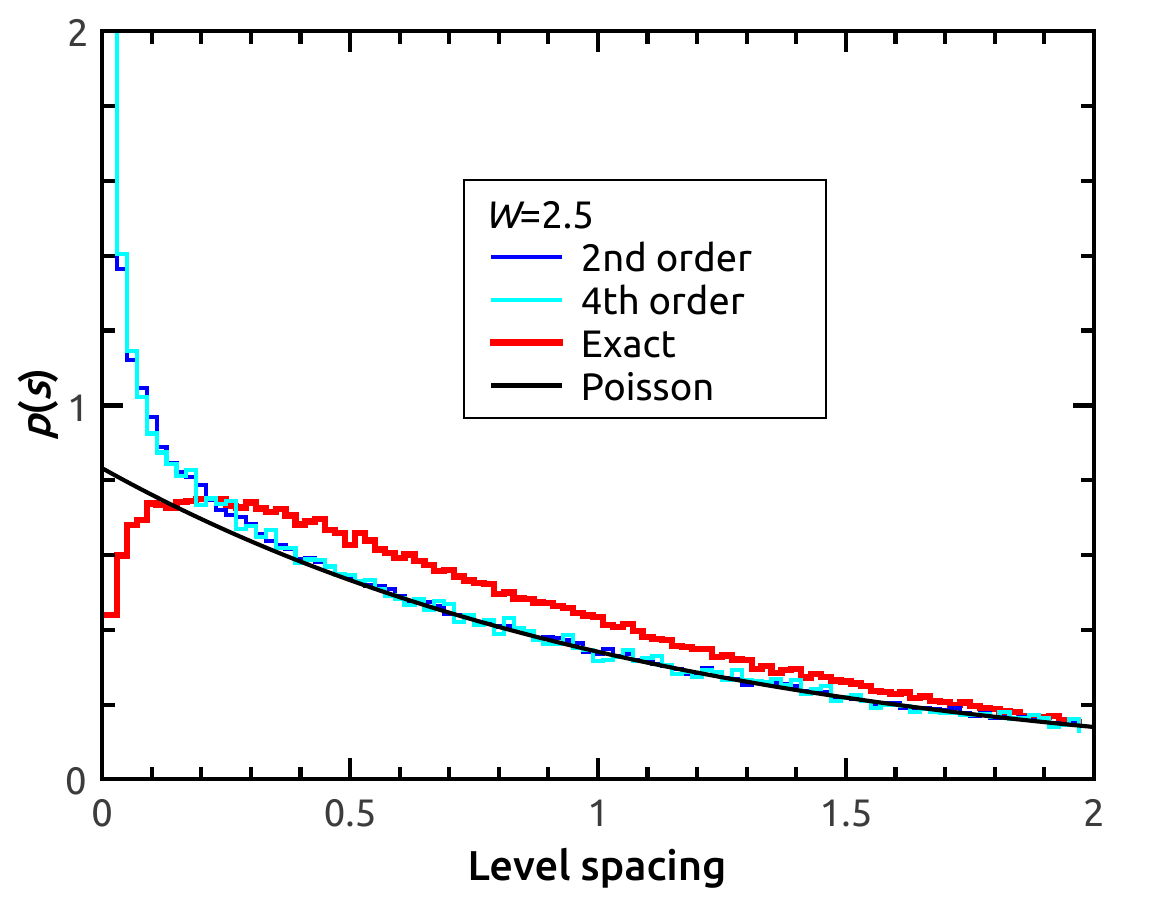}
	\includegraphics[width=.32\textwidth,height=4cm]{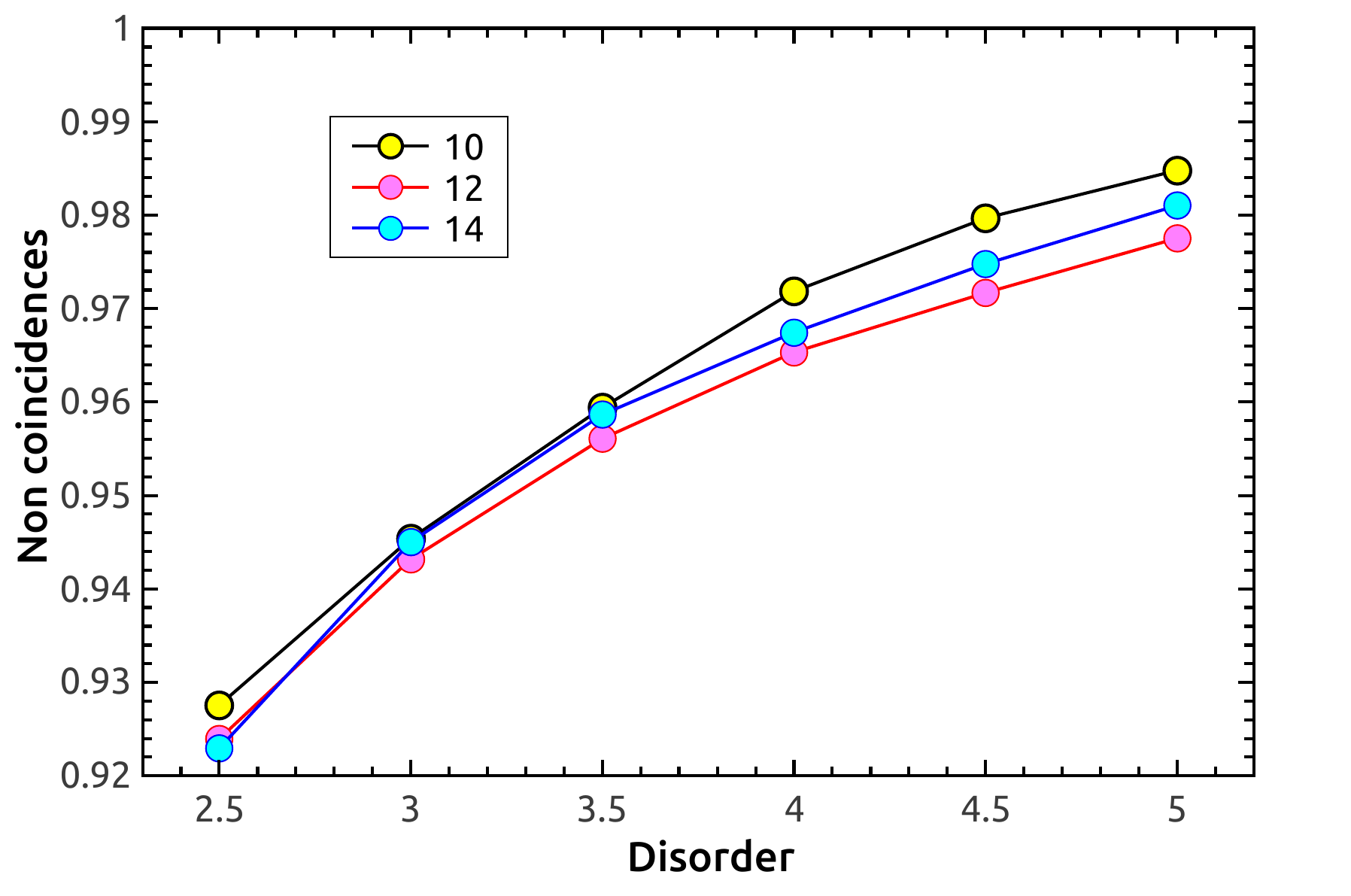}
	\caption{Histogram of the level spacing for our procedure up to second order(blue) and for exact diagonalization (red). The black curve is a fit of our results to a Poisson distribution. The system size is $L=12$ and the range of the disorder is $W=5$ for the left panel and $W=2$ for the central panel.
	Right panel:	Succes ratio of the number of excited states achieved by  our procedure as a function of disorder for system sizes  $L=10$ (black), 12 (red) and 14 (blue).}\label{hist}
\end{figure*}

We have studied the level spacing distribution for  system sizes 10, 12 and 14, for which we can get all states by exact diagonalization and at the same time we can generate states corresponding to all possible combinations of IOM.
We have calculated the level spacing distribution, normalized to the average spacing of the 20 nearest neighbors,  for exact diagonalization results and for our method up to second and up to fourth order. 
The results are presented in Fig.\ \ref{hist}. Panels {\bf a} and {\bf b} correspond to disorders  $W=5$  and 2, respectively.
The red curves correspond to exact results, while the blue curve correspond to our results up to second order with a cut-off $\lambda_{\rm min}=10^{-4}$ and with diagonalization of the projected Hamiltonian.
The black curves in Fig.\ \ref{hist} are a fit of our results to the Poisson distribution.
The exact results show Poisson statistics for  $W=5$ and tend to Wigner-Dyson for  $W=2$, in agreement with previous works \cite{Pal:2010gr} and with the existence of an extended to localized transition for a critical disorder between 2 and 5. 
Our results show Poisson statistics for all disorders considered, as expected since we are  cutting the number of density operators kept in the  Hamiltonian  and all terms are needed in order to reproduce the correct level repulsion.
We observe a peak in our results at very small spacings which we interpret as due to different states collapsing in the same one. The area of this peak is a measure of the success rate. We take the success rate as one minus the area between our curve and the Poisson distribution.
In Fig.\ \ref{hist} {\bf c} we plot the success ratio as a function of disorder for our results up to second order for system sizes $L=10$ (red), 12 (black) and 14 (blue).  Fourth order results are indistinguishable from second order results as long as the level spacing is concerned. 

The succes ratio in Fig.\ \ref{hist} {\bf c}  is rather large, specially in the localized regime, where we only miss less than 4 \% \ of the states.  It is weakly size dependent. To our knowledge, it is the larger success rate achieved in this type of calculations. Obtaining highly excited states of interacting systems is a very   hard task and missing states often results in a biased selection of other states \cite{ratio}. 

\begin{figure}
\includegraphics[width=.45\textwidth]{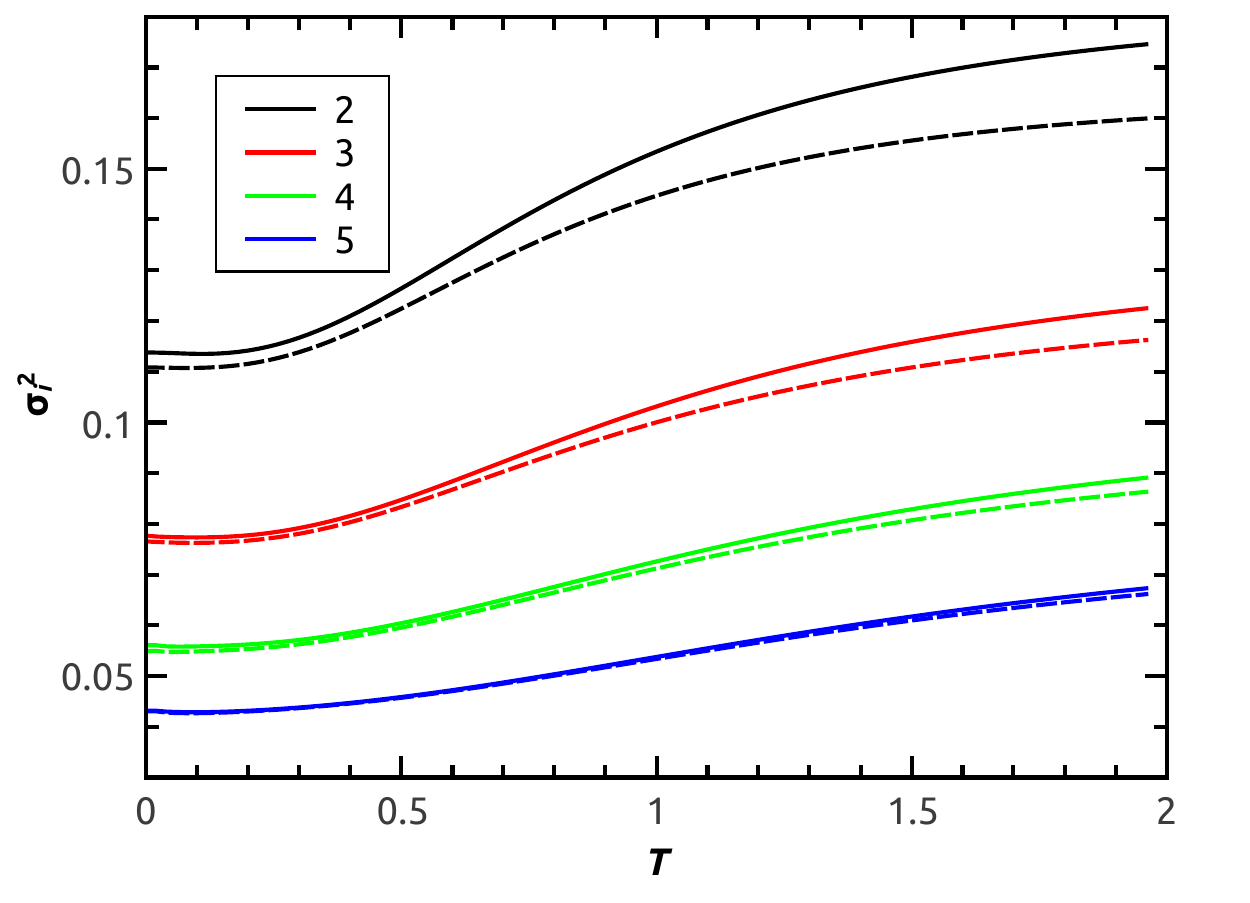}
\caption{Number variance  $\sigma_{L/2}^2$ as a function of temperature for our procedure up to second order (dashed curves) and for exact diagonalization (continuous curves).
The range of disorder is $W=2$ (black), 3 (red), 4 (green), and 5 (blue), and the system size is $L=12$.}
\label{excited}
\end{figure}

Even for moderate size systems, it is not possible to ascertain the precision of a numerical method for excited states from the energy eigenvalues, since there the average spacing is smaller than the error in the energy.
So, we   estimate this precision from the temperature dependence of the variance of the number operator $\sigma_{L/2}^2$, whose quality depends on both the precision in the energy eigenvalues and in the expectation value of the operator.
In Fig.\ \ref{excited} we show $\sigma_{L/2}^2$ as a function of temperature for four values of the disorder and a system size $L=12$. The continuous line corresponds to exact calculations and the dashed line to our procedure up to second order.

We note that our procedure is excellent for all $T$ in the localized regime, blue and green curves, while for lower values of the disorder is good at low $T$, but get worse at hight values of the temperature, where a delocalization transition is  expected.

It is also interesting to know how our procedure behaves for excited states in relatively large systems.
To this aim, we have computed  $\sigma_{L/2}^2$ at infinite temperature, i.e., chosen excited states at random.
In Fig.\ \ref{variance} we plot  $\sigma_{L/2}^2$ at infinite $T$, upper set of curves, as a function of disorder for system sizes between $L=12$ and 30.
The continuous curves are for results from exact diagonalization, for sizes $L=12$ and 16, while symbols, joint by dashed curves, correspond to our results up to second order.
The exact results are almost independent of system size for large disorder, but they increases with size for small disorders, reflecting the many-body delocalization transition around $W_{\rm c}\approx 3.5$.
Our results agree with exact results fairly well in the localized regime, but not so  in the extended region where $\sigma_{L/2}^2$ does not  increase with system size as it should.
We think that this is due to the fact that excited states with very far apart resonances are difficult to select by our method (and probably by most methods) and there is a bias against these more extended states. If we consider transitions up to fourth order and include more excitations in our finite density approach, we should be able to reach this type of states.
It is, in fact surprising that the simplest possible implementation of our procedure, that is, including transitions up to second order only, produces relatively good results in the localized regime.

\section{Outlook}
\label{SecOut}

We have developed a new strategy to perform displacement transformations, by focusing on a specific reference state. Only transformations affecting expectation values for the reference state are included, leading to a higher efficiency. A final diagonalization of the subspace of states up to a fixed number of electron-hole excitations leads to high accuracy results. 

In practice, the displacement transformation are chosen such that they minimize or maximize the energy expectation value with respect to a reference state. This applies equally well to ground states as to excited states. In the latter case, we also minimize the energy variance with respect to the reference state. 

Renormalization group ideas along the lines developed for matrix product states in the so-called density matrix renormalization group can be naturally incorporated into our scheme \cite{White, Thiery}.
One can start diagonalizing with displacement transformations a Hamiltonian for a small system of size $L$ and, at the same time, keep track of how the operators
\begin{equation}
 t c_{L}^\dag c_{L+1}+ U n_L n_{L+1} +{\rm h.c.}
\label{join}\end{equation}
are transformed. These operators are included to increase the system size $L$ with an extra site to $L+1$. 
Once the original system is in diagonal form, we add a new site at $L+1$, that is, we add to the Hamiltonian the transformed operators of Eq.\ (\ref{join}) plus the site energy $\epsilon_{L+1} n_{L+1}$.
The procedure is repeated iteratively.

One can consider a large system right from the beginning, but perform only displacement transformations involving the first site. Once this site enters solely in diagonal terms, one can fix its occupation so that further transformations will not affect it. The same procedure is now applied to site two, and so on.
Alternatively, one can start transforming the whole Hamiltonian and `freeze' sites of either high or low energy along the diagonalization process. With `freezing' we mean substituting the corresponding number operator $n_i$ by either 1 or 0.

We also need to credit an earlier approaches to construct many-body eigenstates beyond the gaussian approximation. 
 Nooijen\cite{Nooijen2000} similarly developed a method to construct many-body states using transformations involving fourth order operators, though this method has not been implemented numerically.

Unlike for matrix product states, our procedure can be directly extended to any number of dimensions or even to amorphous systems with random links. In particular, it should be specially suited for molecular orbital calculations and for nuclear structure determination \cite{Szalay}.
The suitability of the method to deal with excited states should be relevant in the study of Rydberg atoms \cite{Rydberg,Rydberg2} and in
excited states in molecules \cite{Molecules,Rydberg3}.

Displacement transformations can also be used to obtain even and odd normalized zero modes in random interacting Majorana models \cite{Monthus1}.

Finally, we  want to remark that our method is a natural extension of the Hartree-Fock approximation for the construction of non-gaussian states. 
As an extension of Hartree-Fock theory specially adequate in the localized regime, this approach can be relevant in quantum chemical and nuclear matter calculations. We therefore expect that our method can lead to a new quantitive results in many fields.

\acknowledgements
M.O. and A.M.S. were supported by Fundaci\'on S\'eneca grant 19907/GERM/15 and by Spanish MINECO  Grant No. FIS2015-67844-R. L.R. was supported by the SNSF through an Ambizione Grant.

\appendix
\section{Action of the displacement transformations}

A general displacement transformation  is defined through Eq.\ (\ref{displacement}).
A density operator of an state appearing in any creation or annhilation operator in $X$ transforms as
\begin{eqnarray}
\tilde{n}_\delta&=&{\cal D}_X^\dag(\lambda) n_\delta {\cal D}_X(\lambda)\label{density}\\
&=&n_\delta\pm \frac{1}{2}\sin (2\lambda) (X^\dag +X) \mp \sin^2(\lambda) (X^\dag X-XX^\dag)\nonumber
\end{eqnarray}
the upper (lower) sign applies when $\delta$ corresponds to a creation (anhilation) operator in $X$.
The operator $X$ itself transforms as
\begin{equation}
\widetilde{X^\dag +X}
=\cos (2\lambda) (X^\dag +X) - \sin (2\lambda) (X^\dag X-XX^\dag)
\label{quantum}\end{equation}
We can choose $\lambda$ in order to cancel all the terms $X^\dag +X$ appearing in the transformed Hamiltonian.
The sum of the coefficients of all the classical terms with number operators in the quantum part of $X$ is equal to the energy difference $\Delta \epsilon_X$, Eq.\ (\ref{difference}) between the two configurations involved in the transition $X$.
As $V_X$ is the coefficient of the term $X^\dag +X$ in the Hamiltonian, condition Eq.\ (\ref{tan2})
give us the strength of the displacement transformation that cancel all these terms.

It is straight forward to deduce from Eqs.\ (\ref{density}) and (\ref{quantum}) that the condition for canceling the quantum term $X^\dag +X$ is equivalent to maximizing/minimizing the classical term $X^\dag X-XX^\dag$.

 One can in principle diagonalize the interacting Hamiltonian by the application of successive displacement transformations, starting with those involving operators of the form $c_j^\dag c_i$ and continuing with higher (larger number of terms) order operators. The transformation of quantum terms different from $X^\dag +X$ generates other new terms, but they are always of higher order than the original term. To transform away terms of a given order, it is usually convenient to perform the displacement transformations that most reduce the energy first, although other schemes are also possible, for example, in the order of decreasing strength $V_X$.
 Although some times are generated terms with larger strengths, the general trend is a systematic decrease of the maximum remaining strength.

\end{document}